\documentstyle[epsfig,natbib2,natbibmnfix]{mn}

\title[Origin of the Monoceros Ring]{A Sagittarius-Induced Origin
  for the Monoceros Ring}
\author[Michel-Dansac, Abadi, Navarro, Steinmetz]{L\'eo
  Michel-Dansac$^{1}$\thanks{E-mail: leo.michel-dansac@univ-lyon1.fr (LM-D),
    mario@oac.uncor.edu (MGA), jfn@uvic.ca (JFN), msteinmetz@aip.de (MS)},
  Mario G. Abadi$^{2\,\star}$, Julio F. Navarro$^{3\,\star}$, 
  and Matthias Steinmetz$^{4\, \star}$\\
  $^{1}$Centre de Recherche Astrophysique de Lyon, Universit\'e de Lyon,
  Universit\'e Lyon 1, Observatoire de Lyon, \\ Ecole Normale Sup\'erieure de
  Lyon, CNRS, UMR 5574, 9 avenue Charles Andr\'e, Saint-Genis Laval, 69230,
  France\\
  $^{2}$Instituto de Astronom\'ia Te\'orica y Experimental (IATE),
  Observatotio Astron\'omico de C\'ordoba and CONICET, \\ Laprida 854
  X5000BGR C\'ordoba, Argentina\\
  $^{3}$Department of Physics and Astronomy, University of Victoria,
  3800 Finnerty Road, Victoria, BC V8P 5C2, Canada\\
  $^4$Leibniz-Institut f\"ur Astrophysik Potsdam (AIP), An der Sternwarte 16, 14482 Potsdam, Germany
}
\begin{document}

\date{Accepted 2011 February 18. Received 2011 February 17; in original form 2010 December 3}

\pagerange{\pageref{firstpage}--\pageref{lastpage}} \pubyear{2002}

\maketitle

\label{firstpage}

\begin{abstract}
  The Monoceros ring is a collection of stars in nearly-circular
  orbits at roughly 18 kpc from the Galactic center. It may have
  originated (i) as the response of the disc to perturbations excited
  by satellite companions or (ii) from the tidal debris of a disrupted
  dwarf galaxy.  The metallicity of Monoceros stars differs from that
  of disc stars at comparable Galactocentric distances, an observation
  that disfavours the first scenario. On the other hand, circular
  orbits are difficult to accommodate in the tidal-disruption
  scenario, since it requires a satellite which at the time of
  disruption was itself in a nearly circular orbit. Such satellite
  could not have formed at the location of the ring and, given its low
  mass, dynamical friction is unlikely to have played a major role in
  its orbital evolution. We search cosmological simulations for
  low-mass satellites in nearly-circular orbits and find that they
  result, almost invariably, from orbital changes induced by
  collisions with more massive satellites: the radius of the circular
  orbit thus traces the galactocentric distance of the collision.
  Interestingly, the Sagittarius dwarf, one of the most luminous
  satellites of the Milky Way, is in a polar orbit that crosses the
  Galactic plane at roughly the same Galactocentric distance as
  Monoceros. We use idealized simulations to demonstrate that an
  encounter with Sagittarius might well have led to the
  circularization and subsequent tidal demise of the progenitor of the
  Monoceros ring.
\end{abstract}


\section{Introduction}

In a hierarchical universe galaxies are expected to accrete a number
of smaller systems through their lifetime.  These frequent accretion
events leave distinct imprints most easily identified in the outskirts
of a galaxy, where crossing times are long and tidal features can
survive for up to a Hubble time.  Dramatic evidence in support of this
scenario, in the form of tidal tails and recognizable recent and
ancient streams, has been building up steadily in many nearby
galaxies, and especially in the halo of the Milky Way and of the
Andromeda galaxy \citep[see,
e.g.,][]{Ibata1994,Helmi1999,Ibata2003,Belokurov2006,Martinez-Delgado2009,McConnachie2009}.

Among these features, one of the most intriguing is the Monoceros
ring.  It was discovered by \citet{Yanny2003} in the Galactic
anti-center direction using Sloan Digital Sky Survey (SDSS)
data. Numerous surveys, follow-up observations, and spectroscopic
studies have now shown Monoceros to be a dynamically-coherent,
kinematically-cold, low-metallicity, ring-like stellar structure
spanning $\sim 180^{\circ}$ in Galactic longitude
\citep{Ibata2003,Crane2003,Rocha-Pinto2003,Martin2005,Conn2005b,Martin2006,Conn2007,Conn2008,Juric2008,Ivezic2008,deJong2010,Sollima2011}.

The origin of the Monoceros ring (or ``arc'', since it isn't clear yet
whether the structure persists around the whole Galaxy) is still
controversial. A popular hypothesis ascribes it to debris from a
tidally disrupted satellite galaxy
\citep{Helmi2003b,Ibata2003,Conn2005a,Penarrubia2005}. However, it has also
been argued that it could result from a perturbation to the disc,
possibly linked to fly-by encounters with massive substructures or
satellite companions \citep{Kazantzidis2008,Younger2008}.

The position, velocity, and metallicity of stars in the ring may be
used to assess the viability of these scenarios.  \citet{Ivezic2008},
for example, report that Monoceros stars have metallicities distinct
from either the halo or the disc at similar Galactocentric distance, a
result that clearly favours the tidal debris hypothesis.
\citet{Penarrubia2005} are able to fit most available data assuming that
the ring originates from the disruption of a satellite that was in a
prograde, almost coplanar, and nearly circular orbit prior to
disruption. 

The main challenge for the latter scenario is to explain how a
low-mass satellite found its way to a nearly circular orbit at a
Galactocentric distance of only $\sim 18$~kpc: given its eventual
disruption, the satellite clearly could not have formed
there. Cosmological simulations show that most satellites are accreted
in highly-eccentric orbits, which means that a mechanism that allows
the satellite's orbit to migrate and circularize must be postulated.
The low mass of Monoceros' suspected progenitor (estimated to be $\sim
6 \times 10^{8} \, M_{\odot}$ by \citealt{Penarrubia2005}) excludes the most obvious
possibility: dynamical friction with the Galaxy's dark halo.

We present in this Letter a mechanism capable of circularizing
efficiently the highly-eccentric orbit of a low-mass satellite
(Sec.~\ref{sec:OrbCirc}). We then study how this mechanism applies to
the formation of the Monoceros ring (Sec.~\ref{sec:MonForm}). We end
with a brief summary in Sec.~\ref{sec:conc}.

\begin{figure}
  \includegraphics[width=84mm]{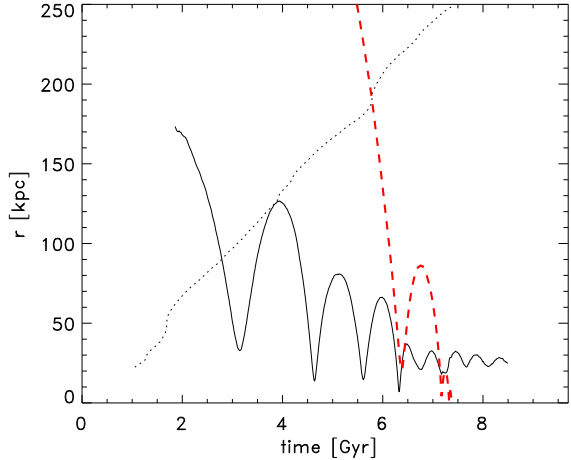}
  \caption{Evolution of the galactocentric distance of two
    satellites selected from a galaxy formation simulation in the LCDM
    cosmogony \citep{Abadi2003a}. The satellites are the same as those
    whose orbital tracks are shown in Figure~\ref{fig:xyz}. The orbit of the less
    massive satellite (shown with solid black line) changes abruptly
    as a result of a close encounter with the more massive satellite
    (shown with thick dashed red line) at $t\sim 6.3$ Gyr. The dotted
    line shows the virial radius of the main galaxy. The more massive
    satellite is disrupted soon after it is accreted, during its
    second pericentric passage. The less massive one is left on a
    nearly circular orbit and is able to complete $4$ to $5$
    revolutions before being fully disrupted.}
    \label{fig:rvst}
\end{figure}

\begin{figure}
  \includegraphics[width=84mm]{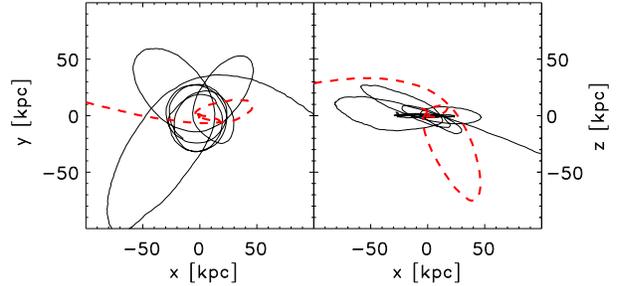}
  \caption{Orbital tracks (left: ``face-on'' view; right: ``edge-on''
    view) of the two satellites shown in Fig.~\ref{fig:rvst}.  The
    orbit of the less masive satellite (shown with solid black line)
    circularizes abruptly after colliding with the more massive
    system. The main galaxy (not shown) is at the coordinate origin at
    all times. }
 \label{fig:xyz}
\end{figure}

\section{Orbital Circularization through 3-body encounters}
\label{sec:OrbCirc}

We have used the N-body/gasdynamical galaxy formation simulations
reported by \citet{Steinmetz2002} and then analyzed in detail by
\citet{Abadi2003a,Abadi2003b,Meza2005} and \citet{Abadi2006}, to search for the
presence of low-mass satellites (masses not exceeding $\sim 1\%$ of
that of the main system) in nearly circular orbits. (Interested
readers may wish to consult those papers for technical details.) These
satellites are then traced back in time in order to recover their full
trajectories.

Although the eight simulations we analyze are of relatively low
resolution (about $10^{5}$ particles per galaxy), we are able to find
a few examples of satellites whose orbital eccentricity does not
exceed $e\equiv (r_{\rm apo}-r_{\rm per})/(r_{\rm apo}+r_{\rm
  per})\sim 0.2$ at the time of their disruption or at $z=0$.  Most
of these satellites were actually accreted into the halo of the main
galaxy in highly-eccentric orbits, only to see their orbits abruptly
become almost circular later on due to complex orbital changes
  that occur during first accretion or to collisions with more massive
  substructures. Collisions are more effective at modifying the orbit
  of satellites far from the center, given the reduced force exerted
  by the main Galaxy at such distances. This may have interesting
  implications for the puzzling low-eccentricity of the orbit of the
  Fornax dwarf spheroidal \citep{Piatek2007}. Although collisions seem
to be the main mechanism for orbital circularization of low-mass
satellites, we note that our simulation series is small, and that this
conclusion should therefore be regarded as tentative until confirmed.

We show an example of the orbital evolution of such satellites in
Figs.~\ref{fig:rvst} and ~\ref{fig:xyz}. These figures show the
evolution of the galactocentric distance and the orbital tracks of two
satellites accreted by the main galaxy, respectively.  The less
massive of the two is accreted early, crossing the virial radius of
the main galaxy (shown with a dotted line in Fig.~\ref{fig:rvst}) for
the first time at $t\sim 2.75$ Gyr{\footnote{Although we quote times
    below in Gyr for ease of reference to the original simulation, we
    note that the simulation was not meant to reproduce the Milky Way
    and therefore these numbers should be taken only as indicative and
    cannot be compared directly with those of Galactic
    satellites.}}. It is in a highly eccentric orbit, reaching a first
pericentric radius of $\sim 32$ kpc from a turnaround radius exceeding
$150$ kpc. Because of the relatively large pericentric radius and the
brief time the satellite spends near pericenter, it survives for
several orbits, losing gradually its mass. Its apocenter and
pericentric radii both shrink as the main galaxy grows in mass. At
$t\sim 6.3$ Gyr, however, the satellite's orbit changes abruptly as a
result of a close encounter with another, more massive satellite
(shown with a thick dashed line in Fig.~\ref{fig:rvst}). The more
massive satellite disrupts quickly in the tidal field of the main
halo, and does not survive its second pericentric passage. The less
massive satellite, on the other hand, survives $4$-$5$ full
revolutions in a nearly circular orbit ($r_{\rm apo}\sim 30$ kpc, $r_{\rm
  per}\sim22$ kpc, $e\sim0.15$) before also disrupting fully.

Fig.~\ref{fig:xyz} shows two different orthogonal projections of the
orbital tracks of the two satellites. Note that the orbital planes of
the two satellites are almost at right angles, and that the radius of
the circular orbit into which the less massive satellite settles
coincides with the galactocentric distance of the encounter between
the two at $t=6.3$ Gyr. Circularization requires both the angular
  momentum and energy of the orbit to be reduced by the
  encounter. This is most easily achieved if the closest approach
  between satellite and massive perturber happens when the perturber
  crosses the orbital plane of the satellite. This is the time when
  the perturbing force is greatest; the coplanar arrangement ensures
  that the direction of the perturbing torque is coincident with the
  satellite's orbital angular momentum, maximizing the effect on the
  orbit.  We have carefully reviewed the interaction shown in
Fig.~\ref{fig:xyz} to confirm that the orbital change of the less
massive system is indeed almost exclusively due to the acceleration
and torque exerted by the more massive satellite during the collision.

This provides a prime example of the way in which a low-mass satellite
can have its orbit circularized before disruption, even when dynamical
friction effects are negligible. Although clearly there is an element
of chance here (i.e., not all close encounters between satellites lead
to circularization) the fact that we have found a few examples of this
mechanism at work in the eight simulations we have analyzed leads us
to conclude that one should not dismiss the possibility that this
process may apply to Monoceros' progenitor. To strengthen the case, we
need (i) to identify the culprit, and (ii) to verify that the
collision could have taken place at the same Galactocentric distance
of Monoceros.  We explore these issues below.

\begin{figure}
  \includegraphics[width=84mm]{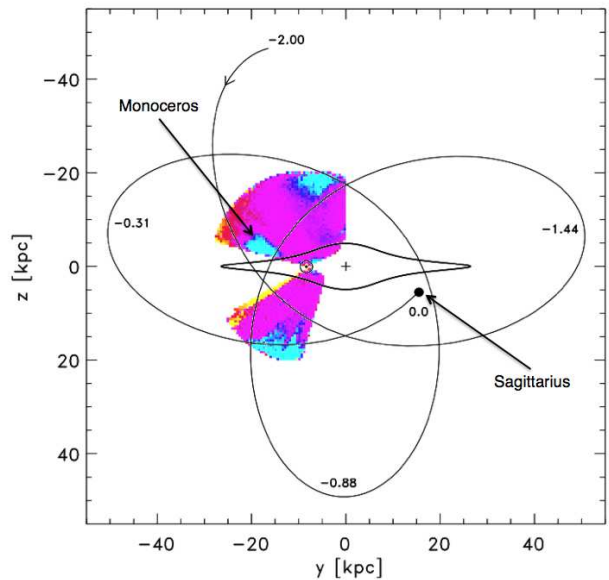}
  \caption{Orbital track of the Sagittarius dwarf during the last
    2~Gyr, projected on a polar plane that contains the Sun. An arrow
    indicates the direction of rotation, and small numbers label the
    lookback time since the various apocentric passages. In the
    footprint of the Galactic disc the cross indicates the Galactic
    centre, and a dotted circle the position of the Sun. The filled
    circle marks the present-day position of Sagittarius. The color
    contours show the stellar overdensities identified by
    \citet{Juric2008} in the SDSS volume, taken from the bottom-left
    panel of their Fig.~23. The Monoceros ring is easily identified in
    this projection, and is indicated by an arrow.}
  \label{fig:sag}
\end{figure}

\section{Application to Monoceros}
\label{sec:MonForm}

Fig.~\ref{fig:sag} shows the (past) orbital track of the Sagittarius
dwarf (solid line), shown projected on a plane perpendicular to the
Galactic disc that contains the Sun. The track follows the most recent
$2$ Gyr of the orbit, assuming a non-evolving, rigid potential for the
Milky Way that consists of a spherical dark matter halo and a stellar
disc. The halo follows a Hernquist profile \citep{Hernquist1990} of
mass $M_{\rm H}=1.4\times10^{12}$~M$_{\odot}$ and characteristic
radius $r_{\rm H}=20$~kpc. The disc potential is modeled by a
Miyamoto-Nagai disc of mass $M_{\rm MN} = 6\times10^{10}$~M$_{\odot}$ and
scale-lengths $a_{\rm MN}=4$~kpc and $b_{\rm MN}=1$~kpc.

The footprint of the Galactic disc is shown in Fig.~\ref{fig:sag} to
guide the eye; the position of the Sun is shown with a dotted
circle. Fig.~\ref{fig:sag} also shows, in color, the residuals from a
smooth Milky Way model obtained by \citet{Juric2008} from SDSS data
and projected azimuthally onto the same plane. These residuals show
the location of several notable stellar streams/overdensities within
the SDSS volume: the one just above and to the left of the Sun is the
Monoceros ring, seen ``edge-on'' in this projection.

Inspection of Fig.~\ref{fig:sag} shows that the nearly polar orbit of
Sagittarius crossed the disc {\it three times} at the same
Galactocentric distance of Monoceros; $\sim 0.6$, $\sim 1.1$, and
$\sim 1.7$ Gyr ago.  This implies that Sagittarius satisfies the first
condition laid out in the previous section in order to link
Sagittarius to Monoceros. The case for such identification is
strengthened by the fact that Sagittarius is one of the most luminous
(and likely most massive) satellites of the Milky Way, and therefore
could have perturbed the orbit of a lesser satellite without itself
experiencing much tidal damage or orbital evolution.

Is it possible that a collision with Sagittarius might have modified
the orbit of a small satellite and placed it onto a circular orbit at
$18$ kpc from the Galactic center? What was the orbit of Monoceros'
progenitor like before the collision?  We explore these issues by
integrating the orbits of test particles in the potential of the
Galaxy.  We place the test particles in coplanar, nearly circular
orbits at $R=18$ kpc, and integrate them {\it backwards} in time,
scanning the results for particles whose orbits change drastically
after collisions with Sagittarius. Reckoned forward in time, these
indicate candidate orbits for possible Monoceros progenitors. We seek
in particular orbits with large initial apocenters (i.e., highly
eccentric) since those are most easily reconciled with cosmological
accretion. In addition, satellites on such orbits would spend little
time in regions of large tidal forces, which would enable the
satellite to survive disruption for several orbital periods
\citep{Penarrubia2008}.

Orbital changes induced by Sagittarius depend mainly on the depth of
its gravitational potential, which scales directly with mass and
inversely with size: the more massive and centrally concentrated
Sagittarius is, the easier it can perturb the orbits of low-mass
satellites.  The simulations were run with {\tt Gadget-2}
\citep{Springel2005b} and model Sagittarius as a rigid spline sphere
\citep{Monaghan1985} of fixed mass and size. Given the uncertainties
in these parameters, we explore a range of them in order to identify
parameter choices that enable the efficient circularization of
low-mass satellites.  We have experimented with $4$ choices for
Sagittarius' mass and size (defined as the radius where the density of
the spline drops to zero): $M_{\rm Sag}=2\times 10^9$, $1 \times
10^{10}$, $2\times 10^{10}$ and $5\times 10^{10} \, M_{\odot}$, as
well as $r_{\rm Sag}=2.8$, $5.6$, $11.2$, and $22.4$ kpc.

\begin{figure*}
  \begin{center}
  \includegraphics[width=150mm]{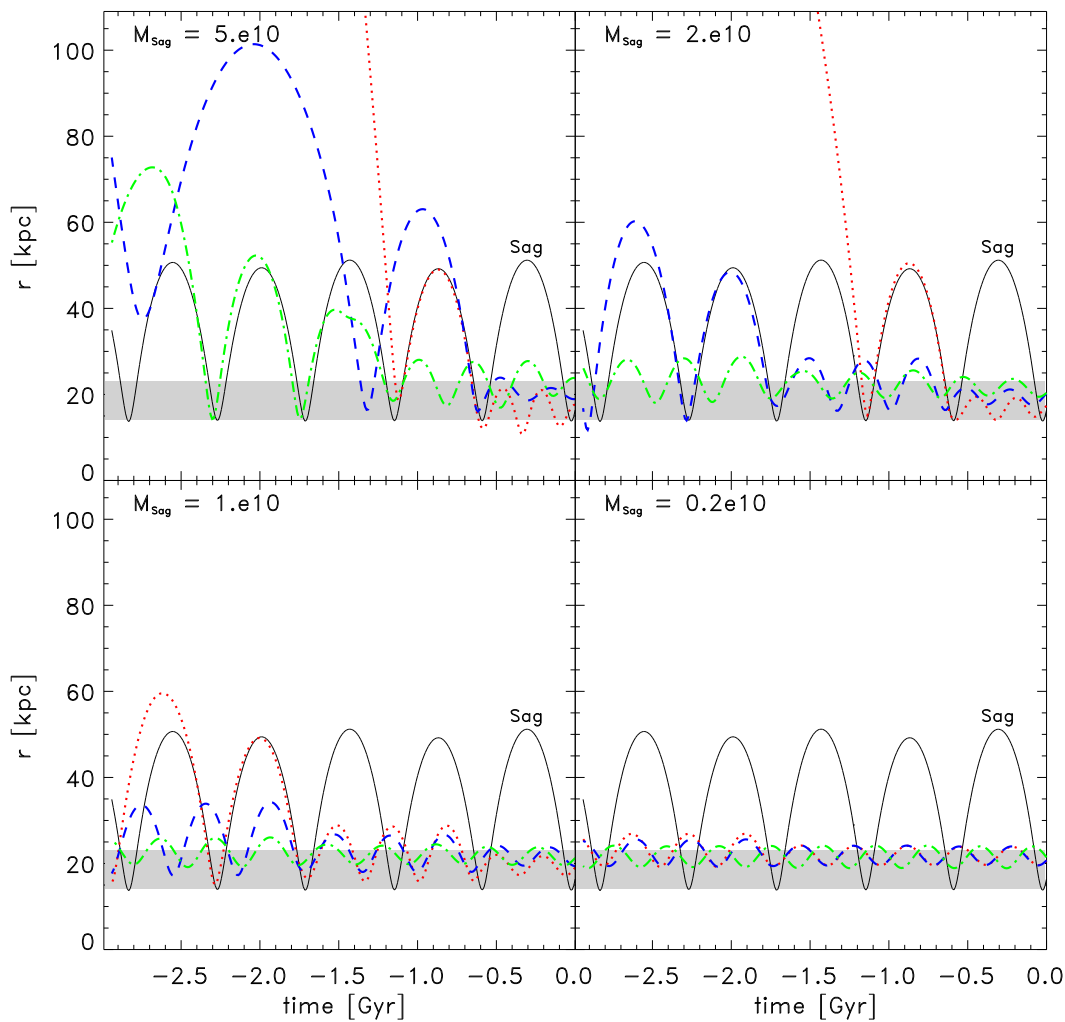}
  \caption{Test-particle orbits in simulations that include the
    effects of Sagittarius in a standard Milky-Way potential. Each
    panel corresponds to different assumptions for the mass of
   Sagittarius (as labeled), which is modeled as a rigid spline
    sphere. Different curves correspond to various choices for
    Sagittarius' size: $r_{\rm Sag}=22.4$ kpc (green
    dot-dashed curve); $5.6$ kpc (blue dashed line); and $2.8$ kpc (red
    dotted curve). The particles shown are chosen amongst those that
    suffer large orbital changes as a result of collisions with
    Sagittarius at a Galactocentric distance of $\sim 18$ kpc. The
    grey band indicated the Galactocentric distance of the Monoceros
    ring.}
 \label{fig:orbits}
  \end{center}
\end{figure*}

Since we are mainly interested in a proof of principle, we show in
Fig.~\ref{fig:orbits} the orbit of one test particle for each choice
of the mass and size of Sagittarius (chosen from among those that have
experienced large orbital changes). Each panel in this figure
corresponds to different Sagittarius masses; the various line types
correspond to different choices for the size, 
as specified in the caption. Thin solid lines outline the orbit
assumed for Sagittarius; the grey band indicates the Galactocentric
distance of Monoceros.

Fig.~\ref{fig:orbits} makes clear that, provided that Sagittarius has
a total mass of at least $10^{10} \, M_{\odot}$, it is able to modify
substantially the orbit of potential Monoceros progenitors. As
expected, increasing $M_{\rm Sag}$ or reducing $r_{\rm Sag}$ induce
more drastic changes in the orbits, allowing  more eccentric
initial orbits to circularize after colliding with Sagittarius. For example, for $M_{\rm
  Sag}=2\times 10^{10} \, M_{\odot}$ and $r_{\rm Sag}=5.6$ kpc, we
find that an orbit with an initial apocenter of $60$ kpc and
pericenter of $15$ kpc can be transformed into a nearly circular orbit
after colliding with Sagittarius.

Interestingly, some of the largest changes are seen to occur after a
couple collisions with Sagittarius, which results from the fact,
alluded to above, that Sagittarius crosses several times the Galactic
plane at about the same Galactocentric distance in the past $2$ Gyr.
In some cases even ``unbound'' orbits may be captured; see, for
example, the dotted curves in the two top panels of
Fig.~\ref{fig:orbits}. Although these are unrealistic for Monoceros,
they are useful as demonstration that, if massive enough, Sagittarius
is capable of effecting substantial changes on the orbits of low-mass
satellites at the Galactocentric distance of the Monoceros ring.

The sensitivity of the results to Sagittarius' potential depth may be
traced to the short time span of the collision. Indeed, when
Sagittarius crosses the Galactic plane at $\sim 18$ kpc it is close to
the pericenter of its orbit and therefore its speed is high, implying
that there is little time for Sagittarius and Monoceros' progenitor to
interact. Had the collision occurred near Sagittarius' apocenter much
lower $M_{\rm Sag}$ would have sufficed to obtain similar changes to
the ones shown in Fig.~\ref{fig:orbits}.

\section{Summary and Discussion}
\label{sec:conc}

We have used numerical simulations to explore a possible origin of the
Monoceros ring. As in earlier work \citep[see,
e.g.,][]{Penarrubia2005}, this scenario envisions Monoceros as the
result of the tidal disruption of a low-mass satellite, but suggests a
compelling explanation for the nearly circular orbit required for the
satellite at the time of disruption. This is important, for the
progenitor clearly could not have formed at its disruption
radius. Furthermore, since circular orbits are extremely rare in
cosmological simulations it is necessary to find a mechanism able
to place a low-mass satellite in a nearly circular orbit prior to
disruption.

We find, using a set of cosmological simulations of galaxy formation,
that the majority of low-mass satellites in such orbits are the
consequence of chance encounters {\it between} satellites. We show an
example that illustrates two requirements of this scenario: the
collision must involve a more massive partner, and it must occur at
the same galactocentric distance as the circularized orbit radius.

These results make the Sagittarius dwarf a potential culprit: it is
one of the most massive satellites of the Milky Way (except for the
Magellanic Clouds) and it has, in the past $2$ Gyr, crossed three
times the Galactic plane at about the same distance as Monoceros. A
series of idealized simulations show that, in order for Sagittarius to
effect substantial changes on satellites that may cross its path at
$\sim 18$ kpc from the Galactic center (the location of Monoceros),
Sagittarius must have been fairly massive at the time when the
collision took place, at least $\sim 10^{10}\, M_{\odot}$. Although
this seems high, it can not be easily ruled out, as recently argued by
\citet{Niederste-Ostholt2010}.

Further modeling should help define and refine observational
  tests of the scenario we propose here. The properties of the ring at
  Galactic longitudes well away from the anti-Galactic center
  direction where it was discovered should be particularly telling,
  and likely to yield diagnostics of the origin of Monoceros. Further
  observational work should aim to secure not only better constraints
  on the total mass of Sagittarius \citep{Penarrubia2011}, but also
  on the kinematics of Monoceros stars. Indeed, much of the work
  presented here is predicated on indications that Monoceros stars
follow nearly circular orbits around the Galaxy. However, the samples
of stars with measured kinematics are small, and contamination by disk
stars may in principle confuse the interpretation. A dedicated
observational campaign design to survey Monoceros around the Galaxy
and to map its kinematics seems essential in order to ascertain the
true origin of this puzzling structure.

\section*{Acknowledgments}

We thank the anonymous referee for a constructive report. 
LMD acknowledges support from the Agence Nationale de la Recherche
(ANR-08-BLAN-0274-01).  JFN thankfully acknowledges support from the
Canadian Institute for Advanced Research.
We also acknowledge computing resources from the CC-IN2P3 Computing  Center 
(Lyon/Villeurbanne - France), a partnership between CNRS/IN2P3 and 
CEA/DSM/Irfu.

\bibliographystyle{mn2e}
\bibliography{ms}

\label{lastpage}
\end{document}